\documentclass[modern]{aastex631}

\newcommand{\teff}{$T_{\text{eff}}$}
\newcommand{\logg}{$\log{(g)}$}
\newcommand{\vsini}{$v\sin{(i)}$}
\newcommand{\xh}[1]{[#1/H]}
\shorttitle{JWST Chemical Abundances}
\shortauthors{Polanski et al.}
\graphicspath{{./}{figures/}}
\begin{document}

\title{Chemical Abundances for 25 JWST Exoplanet Host Stars with \texttt{KeckSpec}}

\correspondingauthor{Alex S.\ Polanski}
\email{aspolanski@ku.edu}

\author[0000-0001-7047-8681]{Alex S. Polanski} 
\affil{Department of Physics and Astronomy, University of Kansas, Lawrence, KS 66045, USA}

\author{Ian J.~M.\ Crossfield}
\affil{Department of Physics and Astronomy, University of Kansas, Lawrence, KS 66045, USA}

\author[0000-0001-8638-0320]{Andrew W. Howard}
\affiliation{California Institute of Technology, Pasadena, CA 91125, USA}

\author[0000-0002-0531-1073]{Howard Isaacson}
\affiliation{501 Campbell Hall, University of California at Berkeley, Berkeley, CA 94720, USA}

\affiliation{Centre for Astrophysics, University of Southern Queensland, Toowoomba, QLD, Australia}

\author[0000-0002-7670-670X]{Malena Rice}

\altaffiliation{51 Pegasi b Fellow}

\affiliation{Department of Physics and Kavli Institute for Astrophysics and Space Research, Massachusetts Institute of Technology, Cambridge, MA 02139, USA}

\affiliation{Department of Astronomy, Yale University, New Haven, CT 06511, USA}

\submitjournal{Research Notes of the AAS}

\begin{abstract}
Using a data-driven machine learning tool we report \teff, \logg, \vsini, and elemental abundances for 15 elements (C, N, O, Na, Mg, Al, Si, Ca, Ti, V, Cr, Mn, Fe, Ni, Y) for a sample of 25 exoplanet host stars targeted by JWST’s first year of observations. The chemical diversity of these stars show that, while a number of their companion planets may have formed in a disk with chemistry similar to Solar, some JWST targets likely experienced different disk compositions. This sample is part of a larger forthcoming catalog that will report homogeneous abundances of $\sim$4,500 FGK stars derived from Keck/HIRES optical spectra.
\end{abstract}

\keywords{Abundance ratios (11), Chemical abundances (224), Metallicity (1031), Stellar abundances (1577), Exoplanets (498)}


\section{Introduction} \label{sec:intro}
The rapid rate of exoplanet discovery in the last two decades has been accompanied by a push to better understand the stars that host them. In particular, the effort to chemically characterize exoplanet hosts has led to a number of large catalogs reporting precise abundances of different elements (\citealt{Adibekyan2012}, \citealt{Brewer2016a}, \citealt{Clark2021}). Refractory elements (Si, Fe, Mg) have especially been of interest due to their potential to break degeneracies in planet composition that arise from knowledge of planet mass and radius alone. The ratios Fe/Si and Mg/Si can place constraints on the size of a rocky planet’s core or the composition of its mantle (\citealt{Dorn2017}, \citealt{Unterborn2017}), while volatile abundances (C, N, O) enable comparisons of stellar to planetary C/O and N/O which can hint at the formation locations of gaseous planets (\citealt{Oberg2011}, \citealt{Mordasini2016}, \citealt{MacDonald2017}). JWST has already begun to conduct its first observations of the atmospheres of other worlds, making a uniform determination of their host star abundances an important and timely step in the interpretation of those results \citep{Pontoppidan2022}.

\section{Sample \& Methods}

We utilized \texttt{KeckSpec} to determine \teff, \logg, \vsini, and \xh{X} for 15 elements \citep{Rice2020}. \texttt{KeckSpec} is a version of the \texttt{\texttt{Cannon}} \citep{Ness2015} designed for application on spectra from the High Resolution Echelle Spectrometer (HIRES, \citealt{Vogt1994}) on the Keck I telescope. The \texttt{Cannon} is a supervised learning algorithm that identifies correlations between the stellar parameters (“labels”) and the flux at every pixel in the stellar spectrum. A generative model is built using a set of stellar spectra with known stellar labels which can then be applied to determine labels for a new set of spectra. The \texttt{Cannon} has been effectively utilized to determine stellar parameters in a number of other surveys including APOGEE \citep{Clark2021}, RAVE \citep{Casey2017}, and LAMOST \citep{Ho2017}.

\texttt{KeckSpec} was trained on the SPOCS catalog (\citealt{Valenti2005}, \citealt{Brewer2016a}) and therefore can only be applied to stars that are likely to reside in the parameter space spanned by that catalog (\citealt{Rice2020}, Table 3). Using HIRES spectra, we have employed \texttt{KeckSpec} to measure a homogeneous
set of stellar abundances for ~4,500 stars with 4700 $<$ \teff $<$ 6600 K
(Polanski et al., \textit{in prep}). Here we present our results for 25
planet-hosting dwarf stars targeted by JWST’s first year of
observations. While some of our targets have abundances in the original SPOCS catalog we provide values from \texttt{KeckSpec} for completeness. 

As discussed in \citealt{Rice2020}, \texttt{KeckSpec} displays a systematic offset that causes it to overestimate abundance values at low metallicity and underestimate them at high metallicity. This was attributed to a bias in the training sample whose labels were determined with \texttt{Spectroscopy Made Easy}. Using \texttt{KeckSpec} abundances for all the SPOCS catalog stars, we fit and correct for the offsets in each abundance value. This process is described in Polanski et al. (\textit{in prep}), but across our sample the average correction for each element was slight: less than the scatter observed when comparing \texttt{KeckSpec} abundances to those of the SPOCS catalog. In Table \ref{tab:vals} we provide the abundances for 15 elements along with \teff, \logg, \vsini, and signal to noise ratio (SNR/pixel) of the spectrum used. The average SNR of our spectra is $\sim 200$, as determined using the median counts in a region centered on 5500 \AA\footnote{Spectra available on request.}.

\section{Results \& Discussion} 

To assess the performance of \texttt{KeckSpec} on our sample, we compare the values we obtain for \teff, \logg, \vsini, and \xh{Fe} to those found through \texttt{SpecMatch (SM) Synthetic} \citep{Petigura2015}. We find that the scatter in \logg, \vsini, and \xh{Fe} to be 0.11, 1.1, and 0.05, respectively which are comparable to uncertainties returned by \texttt{SM}. For \teff, the scatter is 65K which is nearly half the uncertainty quoted by \texttt{SM} underscoring the accuracy of our parameters.

We also compare the C/O and Mg/Si ratios derived from this work to the ones found by \cite{Kolecki2021}, who derived abundances for a JWST target sample that overlaps ours. Their analysis focused on 9 elements as opposed to our 15 and used a more heterogeneous data set whereas we use exclusively HIRES spectra.  The average deviation between the two samples for Mg/Si is -0.07 which is within our uncertainties. For C/O, the difference is higher at -0.16, however, this disagreement is similar to what was found by \cite{Kolecki2021} when comparing to values from SPOCS and is likely due to the fact that NLTE corrections were not applied in the SPOCS catalog.  

The abundances obtained reveal that the systems targeted by JWST are chemically diverse, with abundances for most elements spanning an order of magnitude. Using abundance values for Ti, Ca, Mg, and Si, we calculate the $\alpha$ enhancement relative to iron and find that all stars in our sample, save HAT-P-26, are chemically consistent with being thin disk members. Mg/Si ratios for our sample range from WASP-121’s 0.85 to HD 182488’s 1.10 with a number of targets having Mg/Si $<$ 1, suggesting that these stars could form ``silicon-rich'' companion worlds. Four targets, HAT-P-14 and WASP-17/18/121, form a distinct population in our sample with both low Mg/Si and C/O relative to the Sun. All of our JWST targets have C/O ratios below 0.8, with the highest being HD 189733 and WASP-69 at $\sim 0.65$ \footnote{Figures visualizing these quantities and a CSV of Table 1 can be found at \url{https://aspolanski.github.io/jwst.html}}. 

Stellar abundances are an important piece of the planet composition puzzle, and comparisons with the atmospheric composition determination already being enabled by JWST in the coming months will be crucial to unveiling their formation histories.

\newpage
\shortauthors{}
\begin{longrotatetable}
\movetabledown=6mm
\begin{deluxetable}{lccccccccccccccccccccc}
\tabletypesize{\tiny}
\label{tab:vals}
\startdata
Name & \teff & \logg  & \vsini   & \xh{C} & \xh{N} & \xh{O} & \xh{Na} & \xh{Mg} & \xh{Al} & \xh{Si} & \xh{Ca}  & \xh{Ti} & \xh{V} & \xh{Cr} & \xh{Mn}&\xh{Fe}          & \xh{Ni}          & \xh{Y}           & C/O              & Mg/Si            & SNR  \\
& $\pm77$ & $\pm0.09$ & $\pm0.9$ & $\pm0.05$ & $\pm0.08$ & $\pm0.07$ & $\pm0.05$ & $\pm0.04$ & $\pm0.04$ & $\pm0.03$ & $\pm0.03$  & $\pm0.04$ & $\pm0.06$ & $\pm0.04$ & $\pm0.05$ & $\pm0.03$          & $\pm0.04$          & $\pm0.08$           &               &             & 
\\
\hline
HD 149026 & 5980 & 4.10    & 5.4     & 0.17    & 0.46    & 0.23    & 0.40      & 0.18     & 0.21     & 0.25     & 0.32      & 0.23     & 0.11    & 0.26     & 0.25     & 0.27     & 0.27     & 0.42    & 0.48 $\pm$0.10  & 0.89 $\pm$0.10  & 304  \\
HD 15337  & 5131 & 4.43   & 0.9     & 0.06    & 0.08    & 0.06    & 0.15     & 0.13     & 0.15     & 0.12     & 0.10        & 0.11     & 0.11    & 0.11     & 0.14     & 0.12     & 0.13     & -0.05   & 0.56 $\pm$0.11 & 1.06 $\pm$0.12 & 220  \\
HD 182488 & 5365 & 4.42   & 2.0     & 0.10     & 0.21    & 0.16    & 0.15     & 0.12     & 0.15     & 0.09     & 0.13      & 0.12     & 0.13    & 0.13     & 0.20      & 0.14     & 0.14     & 0.07    & 0.48 $\pm$0.10  & 1.11 $\pm$0.13 & 265  \\
HD 189733 & 5005 & 4.49   & 1.8     & -0.12   & -0.19   & -0.21   & -0.03    & -0.04    & -0.02    & 0.02     & 0.04      & -0.01    & -0.01   & 0.02     & 0.01     & 0.01     & -0.03    & -0.13   & 0.67 $\pm$0.14 & 0.92 $\pm$0.11 & 328  \\
HD 19467  & 5733 & 4.38   & 1.8     & -0.03   & -0.03   & 0.16    & -0.15    & -0.06    & 0.01     & -0.05    & -0.10       & -0.02    & -0.05   & -0.16    & -0.26    & -0.14    & -0.12    & -0.09   & 0.36 $\pm$0.07 & 1.02 $\pm$0.12 & 302  \\
HD 209458 & 6042 & 4.30    & 3.0     & 0.00     & -0.01   & 0.13    & -0.03    & 0.03     & -0.03    & 0.03     & 0.08      & 0.06     & 0.04    & 0.05     & -0.04    & 0.05     & 0.01     & 0.06    & 0.42 $\pm$0.08 & 1.05 $\pm$0.12 & 258  \\
HD 75732  & 5258 & 4.42   & 0.7     & 0.25    & 0.54    & 0.30     & 0.50      & 0.29     & 0.36     & 0.28     & 0.27       & 0.27     & 0.28    & 0.29     & 0.42     & 0.30      & 0.37     & 0.18    & 0.50 $\pm$0.10   & 1.08 $\pm$0.13 & 320  \\
HD 80606  & 5561 & 4.34   & 2.9     & 0.28    & 0.42    & 0.38    & 0.40      & 0.31     & 0.37     & 0.31     & 0.30        & 0.30      & 0.30     & 0.29     & 0.40      & 0.30      & 0.36     & 0.23    & 0.44 $\pm$0.09 & 1.03 $\pm$0.12 & 263  \\
HD 22046  & 5060 & 4.54   & 1.9     & -0.08   & -0.24   & -0.06   & -0.14    & -0.10     & -0.10     & -0.03    & 0.00        & -0.04    & -0.07   & -0.01    & -0.08    & -0.03    & -0.12    & 0.04    & 0.52 $\pm$0.11 & 0.89 $\pm$0.11 & 278  \\
Kepler-51 & 5665 & 4.67   & 5.5     & -0.07   & -0.17   & -0.10    & -0.12    & -0.11    & -0.15    & -0.05    & 0.03      & -0.02    & -0.05   & -0.01    & -0.08    & -0.02    & -0.14    & 0.06    & 0.59 $\pm$0.12 & 0.92 $\pm$0.11 & 27   \\
HAT-P-14  & 6320 & 3.96   & 6.2     & -0.15   & 0.15    & 0.22    & -0.13    & -0.08    & -0.30     & -0.02    & 0.04       & 0.06     & -0.11   & -0.02    & -0.33    & -0.02    & -0.18    & -0.04   & 0.24 $\pm$0.05 & 0.91 $\pm$0.12 & 208  \\
HAT-P-1   & 5931 & 4.24   & 2.7     & 0.07    & 0.10     & 0.21    & 0.08     & 0.11     & 0.09     & 0.12     & 0.18        & 0.13     & 0.14    & 0.14     & 0.10      & 0.15     & 0.12     & 0.26    & 0.40 $\pm$0.08  & 1.02 $\pm$0.12 & 210  \\
HAT-P-26  & 5011 & 4.39   & 2.1     & 0.11    & -0.10    & 0.24    & 0.02     & 0.10      & 0.16     & 0.10      & 0.07       & 0.11     & 0.07    & 0.01     & -0.03    & 0.02     & 0.03     & -0.21   & 0.41 $\pm$0.08 & 1.06 $\pm$0.12 & 128  \\
TOI-193   & 5392 & 4.36   & 2.5     & 0.19    & 0.30     & 0.22    & 0.33     & 0.23     & 0.29     & 0.24     & 0.22        & 0.22     & 0.22    & 0.22     & 0.33     & 0.24     & 0.29     & 0.08    & 0.52 $\pm$0.11 & 1.03 $\pm$0.13 & 83   \\
TOI-421   & 5281 & 4.46   & 0.9     & -0.01   & -0.12   & 0.07    & -0.06    & 0.01     & 0.04     & -0.01    & -0.02       & 0.02     & 0.02    & -0.03    & -0.08    & -0.03    & -0.04    & -0.03   & 0.45 $\pm$0.09 & 1.09 $\pm$0.11 & 218  \\
WASP-52   & 5089 & 4.46   & 4.1     & 0.16    & 0.29    & 0.13    & 0.42     & 0.19     & 0.23     & 0.26     & 0.32        & 0.22     & 0.23    & 0.26     & 0.37     & 0.29     & 0.28     & 0.12    & 0.59 $\pm$0.12 & 0.90 $\pm$0.10   & 120  \\
WASP-121  & 6335 & 4.17   & 13.5    & 0.04    & 0.54    & 0.42    & 0.16     & 0.15     & -0.03    & 0.24     & 0.25       & 0.22     & 0.01    & 0.23     & 0.17     & 0.24     & 0.17     & 0.29    & 0.23 $\pm$0.05 & 0.85 $\pm$0.12 & 204  \\
WASP-127  & 5845 & 4.22   & 0.3     & -0.12   & -0.38   & -0.01   & -0.28    & -0.15    & -0.16    & -0.15    & -0.10       & -0.08    & -0.10    & -0.16    & -0.37    & -0.16    & -0.23    & -0.17   & 0.43 $\pm$0.09 & 1.04 $\pm$0.11 & 210  \\
WASP-166  & 6018 & 4.29   & 4.3     & 0.15    & 0.27    & 0.28    & 0.23     & 0.17     & 0.16     & 0.21     & 0.23       & 0.18     & 0.19    & 0.23     & 0.26     & 0.23     & 0.22     & 0.29    & 0.41 $\pm$0.08 & 0.97 $\pm$0.10  & 209  \\
WASP-17   & 6490 & 4.08   & 6.4     & -0.29   & 0.09    & 0.19    & -0.08    & -0.06    & -0.36    & 0.00      & 0.05        & 0.07     & -0.06   & -0.02    & -0.31    & -0.00     & -0.15    & -0.08   & 0.18 $\pm$0.04 & 0.91 $\pm$0.10  & 170  \\
WASP-18   & 6220 & 4.10    & 10.2    & -0.01   & 0.22    & 0.37    & 0.08     & 0.05     & -0.07    & 0.12     & 0.21        & 0.15     & 0.06    & 0.14     & -0.01    & 0.15     & 0.03     & 0.19    & 0.23 $\pm$0.05 & 0.88 $\pm$0.12 & 203  \\
WASP-39   & 5463 & 4.41   & 2.1     & -0.04   & -0.06   & 0.04    & -0.02    & -0.00     & 0.02     & -0.01    & -0.01       & -0.00     & 0.00     & -0.01    & -0.03    & -0.01    & -0.03    & 0.02    & 0.46 $\pm$0.09 & 1.06 $\pm$0.13 & 149  \\
WASP-63   & 5536 & 3.93   & 2.9     & 0.14    & 0.12    & 0.13    & 0.15     & 0.16     & 0.19     & 0.14     & 0.25        & 0.20      & 0.16    & 0.22     & 0.27     & 0.24     & 0.27     & 0.26    & 0.56 $\pm$0.11 & 1.09 $\pm$0.10  & 209  \\
WASP-69   & 4876 & 4.50    & 2.6     & 0.27    & 0.38    & 0.18    & 0.68     & 0.33     & 0.39     & 0.40      & 0.44        & 0.32     & 0.31    & 0.35     & 0.50      & 0.41     & 0.45     & 0.21    & 0.67 $\pm$0.14 & 0.88 $\pm$0.12 & 237  \\
WASP-77A  & 5569 & 4.45   & 3.7     & -0.02   & -0.05   & 0.06    & -0.05    & -0.00     & 0.00      & -0.01    & 0.02       & 0.01     & 0.02    & 0.00      & -0.02    & 0.01     & -0.02    & -0.00    & 0.46 $\pm$0.09 & 1.07 $\pm$0.10  & 211 \\
\enddata
\tablecomments{\teff~units: K. \logg~units: [cm s$^{-2}$]. \vsini~units: km s$^{-1}$. \xh{X}~units: dex}
\end{deluxetable}
\end{longrotatetable}

\bibliography{main}{}
\bibliographystyle{aasjournal}

\end{document}